\documentclass[twocolumn]{aastex61}
\usepackage{graphicx}
\usepackage{epstopdf}

\usepackage{color}
\usepackage{soul}

\shorttitle{The massive green valley galaxies in 3D-HST/CANDELS}
\shortauthors{Y. Gu et al.}

\begin{document}

\title{\bf The morphological evolution, AGN fractions, dust content, environments, and downsizing of massive green valley galaxies at $0.5<z<2.5$ in 3D-HST/CANDELS}

\correspondingauthor{Qirong Yuan}
\email{yuanqirong@njnu.edu.cn}

\author{Yizhou Gu}
\affil{Department of Physics and Institute of Theoretical Physics, Nanjing Normal University,
Nanjing  210023, China;  
yuanqirong@njnu.edu.cn, guyizhou00123@msn.cn}

\author{Guanwen Fang}
\altaffiliation{Guanwen Fang and Yizhou Gu contributed equally to this work}
\affil{Institute for Astronomy and History of Science and Technology, Dali University, Dali 671003, China; wen@mail.ustc.edu.cn
} 

\author{Qirong Yuan}
\affil{Department of Physics and Institute of Theoretical Physics, Nanjing Normal University,
Nanjing  210023, China;  
yuanqirong@njnu.edu.cn, guyizhou00123@msn.cn}

\author{Zhenyi Cai}
\affil{CAS Key Laboratory for Research in Galaxies and Cosmology, Department of Astronomy, University of Science and Technology of China, Hefei, Anhui 230026, China}
\affil{School of Astronomy and Space Science, University of Science and Technology of China, Hefei 230026, Peopleʼs Republic of China}
\author{Tao Wang}
\affil{Institute of Astronomy, University of Tokyo, 2-21-1 Osawa, Mitaka, Tokyo 181-0015, Japan}



\begin{abstract}
To explore the evolutionary connection among red, green, and blue galaxy populations, based on a sample of massive ($M_* > 10^{10} M_{\odot}$) galaxies  at $0.5<z<2.5$ in five 3D-$HST$/CANDELS fields, we investigate the dust content, morphologies, structures, active galactic nucleus (AGN) fractions, and environments of these three galaxy populations. Green valley galaxies are found to have intermediate dust attenuation, and reside in the middle of the regions occupied by quiescent and star-forming galaxies in the $UVJ$ diagram.
Compared with blue and red galaxy populations at $z<2$, green galaxies have intermediate compactness and morphological parameters. The above findings seem to favor the scenario that green galaxies are at a transitional phase when star-forming galaxies are being quenched into quiescent status. 
The green galaxies at $z<2$ show the highest AGN fraction, suggesting that AGN feedback may have played an important role in star formation quenching. 
For the massive galaxies at $2<z<2.5$, both red and green galaxies are found to have a similarly higher AGN fraction than the blue ones, which implies that AGN feedback may help to keep quiescence of red galaxies at $z>2$.
A significant environmental difference is found between green and red galaxies at $z<1.5$. 
Green and blue galaxies at $z>0.5$ seem to have similar local density distributions, suggesting that environment quenching is not the major mechanism to cease star formation at $z>0.5$.  The fractions of three populations as functions of mass support a ``downsizing'' quenching picture that the bulk of star formation in more massive galaxies is completed earlier than that of lower-mass galaxies.

\end{abstract}




\section{Introduction} \label{sec:intro}
A bimodal distribution exists in many aspects of galaxies, such as morphology (spiral vs. elliptical), color (blue vs. red), and kinematics (rotation vs. random motion).
Two peaks in the optical color distribution, based on the Sloan Digital Sky Survey (\citealt{York+00}), are revealed in the local universe (\citealt{Strateva+01, Baldry+04}).
The narrow red peak is usually called `red sequence' (RS), containing abundant quiescent galaxies and a small amount of dusty star-forming galaxies (\citealt{B_M+09}).
The extend blue peak is called `blue cloud' (BC), mainly composed of blue star-forming galaxies (\citealt{Kauffmann+03a,Kauffmann+03b,Baldry+04}). 
The color bimodality of galaxies is also found in ultraviolet (\citealt{Salim+07, Wyder+07}) and infrared bands (\citealt{Walker+13, Lee+15}) .
Furthermore, it is proved that the color bimodality exists at $z \sim 2.5$ (\citealt{Brammer+09}) and at even higher redshift (\citealt{Xue+10,Whitaker+11}).

The joint region between RS and BC is called the `green valley' (GV), which is initially proposed and described in a series of papers (\citealt{,Martin+07, Salim+07, Schiminovich+07, Wyder+07}) on the basis of Galaxy Evolution Explorer (GALEX).
Since GV galaxies exactly lie below the star-forming main sequence (SFMS) (\citealt{Salim+07}), they are a great medium within which to study the star-formation quenching and evolution of galaxies.
It has been pointed out that GV galaxies are a distinct population, rather than a simple mixing (\citealt{Wyder+07,Salim+09,Mendez+11}).
Moreover, GV galaxies are thought to be a transitional population when blue galaxies transform into the red and dead (\citealt{Bell+04,Faber+07,Balogh+11}).
The transition timescale cannot be too long ($ < 1$  Gyr); otherwise, the bimodal distribution could not be so prominent in the color-magnitude diagram (\citealt{Faber+07, Martin+07, Balogh+11}).
In addition, galaxies should not follow a single evolutionary track when crossing over the GV.
\cite{Schawinski+14} propose a scenario assuming distinct evolution of cosmic gas supply and gas reservoirs to interpret that late-type galaxies  are generally quenched more slowly than early-type galaxies.
After the cosmic gas supply is shut off, the quenching of late-type galaxies will take over several gigayears to exhaust the remaining gas by secular and/or environmental processes.
In contrast, the  quenching of early-type galaxies happens quickly, with a timescale of $< 250$ Myr, when the gas supply and gas reservoir are instantaneously destroyed by a major merger.
In more detail, \cite{Pandya+17} propose four dominant evolutionary modes for star formation histories: oscillations on the SFMS, slow quenching, quick quenching, and rejuvenation.
Another evolutionary mode at high redshift $z \gtrsim 1$ has been recently discussed by \cite{Mancuso+16}, where the early-type galaxies are formed via an in situ coevolution scenario instead of a morphological transformation from late-type galaxies.

In order to study the properties of the green galaxy population, several selection criteria have been proposed to define GV in previous work, using empirical color cuts in color-mass/magnitude diagrams, such as $U - V$ (\citealt{Brammer+09}), $U - B$ (\citealt{Mendez+11}),
$\rm{NUV}-r$ (\citealt{Wyder+07}), $[3.4\ \mu m] - [12\  \mu m]$ (\citealt{Lee+15}) and in color-color space, such as $\rm{FUV-NUV}$ versus $\,\,\rm{NUV} - [3.6 \mu m]$ (\citealt{Bouquin+15}).
The color $\rm{NUV} - r$ performs better in selection of green galaxies  than  the colors $u - r$  (\citealt{Baldry+04}) and $g - r$ (\citealt{Blanton+03}), because the blackbody radiation spectra of young stellar populations peak in the near-UV (NUV) band, yeilding a more dynamic range and thus separation in the typical colors of RS and BC galaxies. 
However,  the observed or rest-frame colors cannot reflect the intrinsic activity of star formation because of a wide variety of dust attenuation in galaxies. Some dusty star-forming galaxies are likely to be misclassified as red quiescent galaxies owing to severe dust reddening.
To get rid of the blending of dusty galaxies, \cite{Salim+09} and \cite{Mendez+11} identified dusty galaxies by comparing the rest-frame ($\rm{NUV} - R$) color with the specific star formation rate (sSFR).

With the improvement in technology of fitting the spectral energy distributions (SEDs), it is possible to break the degeneracy of star-formation history (SFH) and dust extinction and to reveal the intrinsic colors of galaxies.
\cite{Pan+13} use the dust-corrected $\rm{NUV}-r$ color to select $\sim 2350$ green galaxies in the COSMOS field at $0.2 < z < 1.0 $, where the most suitable extinction curve is selected for the best-fit template  considering that different extinction curves are expected from galaxy to galaxy (\citealt{Ilbert+09}).
\cite{Pandya+17} select GV galaxies in the sSFR $- M_*$ diagram, where the dust-corrected SFR is derived from its SED-based rest-frame NUV luminosity at 2800\AA  \ and corrected by assuming the \cite{Calzetti+00} dust attenuation curve.
\cite{Brammer+09} find that dust-corrected rest-frame $U-V$ color behaves well in separating dusty starburst galaxies and intrinsically red quiescent galaxies, where the extinction correction of rest-frame $U - V$ color is derived from the SED fitting using the \cite{Calzetti+00} extinction law.
Based on the extinction-corrected rest-frame $U-V$ color, \cite{Wang+17} establish a separation criterion to select massive red, green, and blue galaxy populations  in the GOODS-N and GOODS-S fields, which is more self-consistent with our current understanding that the galaxies tend to be redder with increasing stellar mass and cosmic time.

Several studies have revealed the properties of GV galaxies at low and intermediate redshifts ($z<1.5$) to a certain degree. 
From the morphological angle of view, most GV galaxies are found to be bulge-dominated disk galaxies (\citealt{Mendez+11, Salim+14, Bait+17}).
Quantitative studies show that GV galaxies have intermediate distributions of morphological parameters between RS and BC galaxies, such as S\'{e}rsic index ($n$), concentration ($C$), asymmetry ($A$), smoothness ($S$), and bulge-to-total ratio (\citealt{Schiminovich+07, Mendez+11, Pan+13}).
Morphologies or structure parameters of galaxies seem to be related to the intensity of star formation (e.g., \citealt{Cheung+12, Pan+13, Brennan+15, Brennan+17, Powell+17}).
Moreover, \cite{Bait+17} study the dependence of galaxy morphology on star formation and environment in the  local universe, suggesting that morphology strongly correlates with the sSFR, while the environmental effects on morphology and sSFR are weak for local massive galaxies.

A key question is what physical processes cause the cessation of star formation.  Several  mechanisms have been proposed to interpret the quenching process.
Both internal processes (mass quenching) and external processes (environment quenching)  can  lead to the cessation of star formation (\citealt{Peng+10}).
Mergers, which are violent processes, are one possible mechanism responsible for quenching because they can fuel starbursts that rapidly use up gas and/or they can expel gas through shocks generated by supernovae (\citealt{Springel+05,Robertson+06}).
The active galactic nucleus (AGN) phenomenon is more commonly detected in the GV galaxies (\citealt{Nandra+07, Coil+09, Schawinski+10} ),  suggesting that the large amounts of radiation produced by central AGNs can heat (or expel) the gas within a galaxy to quench star formation (\citealt{Bower+06,Tremonti+07}).
Indeed, observations of cavities (or bubbles) at X-ray (or radio) wavelengths supports the picture where AGNs can quench massive galaxies (\citealt{Fabian+12}).
Some mechanisms, such as strangulation (\citealt{Larson+80}), ram pressure stripping (\citealt{G&G+72}), and harassment (\citealt{F&S+81}), are proposed to explain the environmental effects in star formation quenching.
Which mechanism is responsible for  the star-formation quenching  depends on the properties of individual galaxies.

In this paper, by using the extinction-corrected rest-frame $U - V$ color, we construct a large sample of massive ($M_* > 10^{10}M_{\odot}$) galaxies at $0.5 < z < 2.5$ in five fields of 3D-$HST$/CANDELS (\citealt{Grogin+11,Koekemoer+11,Skelton+14}).
The size of our sample is large enough to be divided into 12 subsamples, which correspond to  red, green, and blue galaxy populations at four redshift bins (with $\Delta z = 0.5$ ).
In order to figure out whether green galaxies represent a transitional population between star-forming  and passive galaxies, we present a  quantitative analysis of the following properties for these three galaxy populations: dust attenuation, morphology, structure parameters, AGN fraction, and local environmental density.
Since previous studies mainly focus on the green galaxies at low and intermediate redshifts (e.g., \citealt{Schiminovich+07, Mendez+11, Pan+13}),  in this paper we improve the redshift limit up to $z \sim 2.5$, following other works that also explicitly define and study GV galaxies up to $z \sim 2.5-3$ (e.g., \citealt{Brennan+15, Brennan+17, Pandya+17, Wang+17}). 
We shall focus on the difference in these properties and its evolution with cosmic time for three populations, which will help us understand what physical processes are dominant during star formation quenching. 
We find a general trend that the buildup of the bulge component over time is accompanied by quenching of star formation.
For massive galaxies at $z > 0.5$, AGN feedback may play an important role in star-formation quneching.

The structure of our paper is as follows. 
We describe the data set of 3D-$HST$ and CANDELS programs and sample selection of massive galaxies in section 2.  
Basic properties of massive galaxies are presented in section 3. 
Our results on morphological and structural evolution are presented in section 4 and 5, respectively. 
The influences of AGN fraction and environmental density is discussed in section 6 and 7, respectively.
In section 8, some implications based on our results are discussed. Finally, a summary is given in section 9. Throughout our paper, we adopt the cosmological parameters as follows: $H_0=70\,{\rm km~s}^{-1}\,{\rm Mpc}^{-1}$, $\Omega_m=0.30$, $\Omega_{\Lambda}=0.70$.

\section{Data and Sample Selection} \label{sec:data}
\subsection{Data Reduction}
Our work is based on the high-quality WFC3 and ACS spectroscopy and photometry from the 3D-$HST$ (\citealt{Skelton+14}) and  CANDELS (\citealt{Grogin+11,Koekemoer+11}) programs, covering over  900 $\rm{arcmin}^{2}$ in five separate fields: AEGIS, COSMOS, GOODS-N, GOODS-S, and UDS.
All these five fields have been observed with space-based (i.e., $HST$/WFC3, $HST$/ACS, \emph{Spitzer}) and many ground-based telescopes. With a wealth of public imaging data available from UV to IR band, it becomes possible to build the SEDs over a wide wavelength range for high-redshift galaxies and then to study galaxy populations over most of cosmic history.

Based on multiwavelength photometric data at wavelength 0.3 $-$ 8.0 $\mu$m, including the CANDELS, 3D-$HST$  imaging and other available imaging data, \cite{Skelton+14} present a photometric analysis.
Using the EAZY code (\citealt{Brammer+08}), they derived photometric redshift ($z_{\rm phot}$)  by fitting the SED of  each galaxy with a linear combination of seven galaxy templates. Comparison of photometric redshifts to the  spectroscopic redshifts ($z_{\rm spec}$) from the literature shows a high precision in the $z_{\rm phot}$ estimate: the normalized median absolute deviation of $\Delta z = z_{\rm phot} - z_{\rm spec}$, i.e., $\sigma_{\rm NMAD} = 1.48 \times \mbox{median}[|\Delta z-\mbox{median}(\Delta z)|/(1+z_{\rm spec})]$, are 0.022,  0.007, 0.026, 0.010, and 0.023 for the AEGIS, COSMOS, GOODS-N, GOODS-S, and UDS fields, respectively (\citealt{Skelton+14}).
In this paper, we prefer to take the spectroscopic redshifts if available. Otherwise, we will use the photometric redshifts instead.
The rest-frame $U-V$  colors derived with the EAZY templates  and the best-fitting redshifts are also taken in further analysis.

Additionally, stellar population parameters (e.g., the stellar mass and dust attenuation) have been derived by  \cite{Skelton+14} with the FAST code (\citealt{Kriek+09}) on the basis of the \cite{BC03} (hereafter BC03)  stellar population synthesis (SPS) models  with a \cite{Chabrier+03} initial mass function (IMF) and solar metallicity.
For obtaining a better estimate of dust attenuation ($A_V$), \cite{Wang+17} adopt the \cite{Maraston+05} SPS models to construct galaxy templates.  The contribution of the asymptotic giant branch (AGB) stars is taken into account in the \cite{Maraston+05} models, which is more reasonable for star-forming galaxies at high redshifts. Thereafter, based on the extinction-corrected rest-frame colors $U-V$, the GV is defined by two mass- and redshift-dependent color boundaries in \cite{Wang+17}.

\cite{Wang+17} have only considered the GOODS-S and GOODS-N fields. In order to ensure systematic consistency on the definition of the GV,
for all five 3D-$HST$/CANDELS fields, we perform the FAST code to reestimate the stellar mass ($M_*$) and dust  attenuation ($A_V$),  following \cite{Wang+17}.
Taking the SPS models of \cite{Maraston+05} with a \cite{Kroupa+01} IMF and solar metallicity, assuming exponentially declining SFHs with the $e$-folding time ranging from $10^8$ to $10^{10}$ yr, we build the galaxy templates with the \cite{Calzetti+00} reddening law and allow dust extinction ($A_V$) to vary from 0 to 4. For avoiding the contamination of polycyclic aromatic
hydrocarbon (PAH) and AGN emission, two longer IRAC bands are excluded in our SED fitting. It has been proved that AGN contamination does not affect the integrated colors of host galaxies significantly (\citealt{Pierce+10, Wang+17}).  \cite{Wang+17} assessed AGN contamination of host-galaxy colors and show that the median difference in rest-frame $U-V$ color between AGN hosts and non-AGN galaxies is less than 0.020 mag. After excluding the two longer IRAC bands, it is not necessary to include the AGN templates in our SED fittings.

Compared with the values of stellar mass and $A_V$  derived with the BC03 models and the \cite{Chabrier+03} IMF,  the average values of our estimates are 0.05 dex lower in mass and 0.18 dex lower in $A_V$.
As described in \cite{Wang+17}, the reason is that our SED fittings use the \cite{Maraston+05} models, which add the contribution of AGB stars.

\subsection{Sample Selection} \label{sec:sample}

\begin{figure*}
\centering
\includegraphics[scale=0.56]{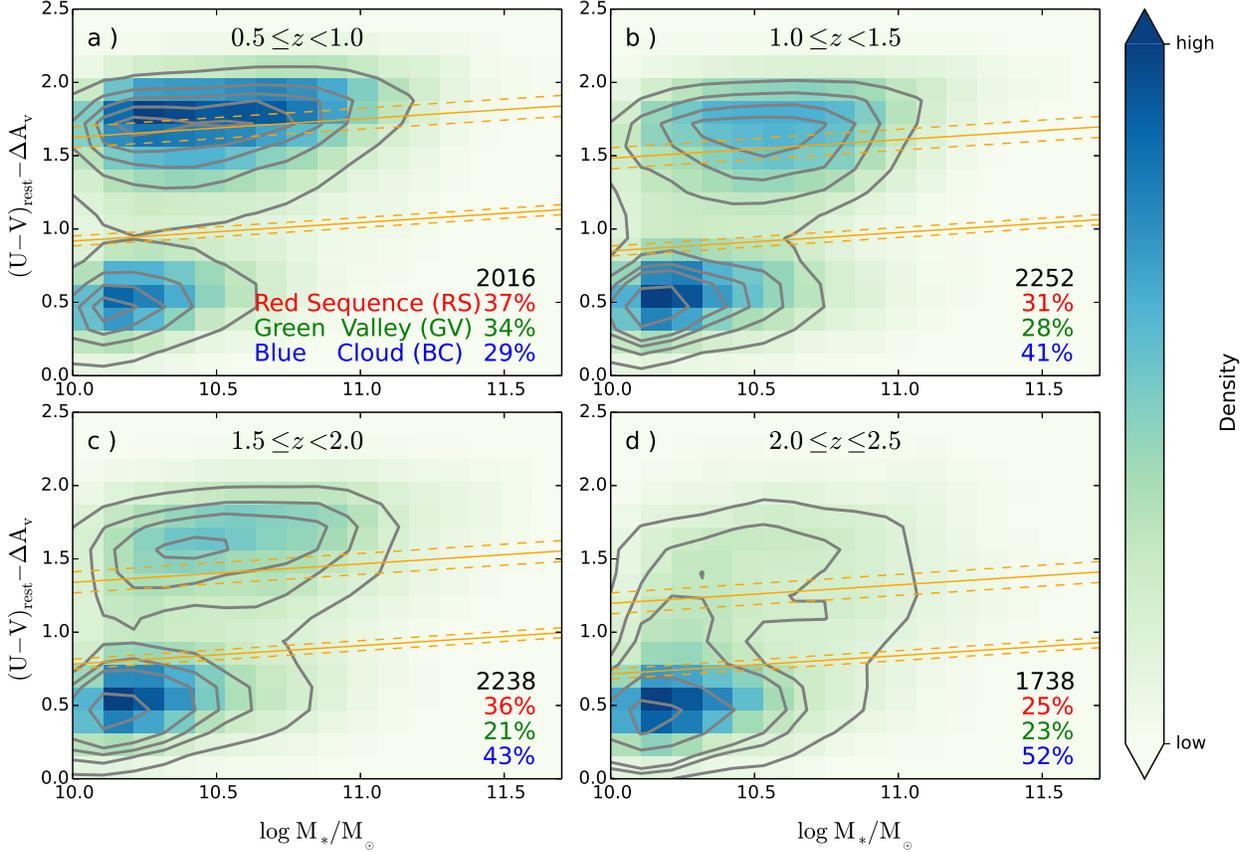}
\caption{Extinction-corrected rest-frame $U-V$ color as a function of stellar mass in four redshift bins.
The contours and  color blocks represent the relative density  in the color-mass relation.
The criteria for each redshift bin, given by \cite{Wang+17}, are shown as orange solid lines (for the mean redshifts) and dashed lines (for each redshift bin).
At the bottom right of each panel,  the subsample size  and proportions of RS, GV, and BC galaxies are presented.
}
\label{fig1}
\end{figure*}

We only choose the galaxies with $\tt use\_phot = 1$  (\citealt{Skelton+14}) in our sample selection, which means that the source
(1) is not a star, or bright enough to be recognized as a galaxy;
(2) is not close to a bright star;
(3)  is well exposed, namely, it requires that each securely detected object be covered in at least two individual exposures in each of the two bands F125W and F160W;
(4) has a signal-to-noise ratio $\rm S/N > 3$ in the F160W images; 
and (5) has a passable photometric redshift fit and a ``non-catastrophic'' stellar population fit, which means $ \log(M_{*}) > 0$.
All  massive galaxies with $\log (M_*/M_{\odot}) > 10$ and $0.5\leq z \leq 2.5$ in five 3D-$HST$/CANDELS fields are selected. This mass threshold ensures a completeness of above 90\% up to $z \sim 2.5$  (\citealt{Grogin+11, Wuyts+11, Newman+12, Barro+13,Pandya+17}), even for the quiescent galaxies which is harder to be detected than the star-forming galaxies at $z<2.5$.
As a result, there are 8244 massive galaxies at $0.5\leq z \leq 2.5$ in our sample,  including $\sim16\%$ galaxies with known spectroscopic redshifts. In order to observe the cosmic evolution of galaxy properties, the sample is divided  into four redshift bins: (i) $0.5 \leq z < 1.0$, (ii) $1.0 \leq z < 1.5$, (iii) $1.5 \leq z < 2.0$, and (iv) $2.0 \leq z \leq 2.5$.

The rest-frame color,  $(U-V)_{\rm rest}$,  stretches over the $\rm 4000\AA$  break, and it can be used to separate passive galaxies from star-forming galaxies. \cite{Bell+04} find the bimodality of  $(U-V)_{\rm rest}$ at $z < 1.0$ and then define a dividing line to separate blue and red galaxies.
However, as mentioned above,  some dusty star-forming galaxies are blended with quiescent RS galaxies in the color$-$mass diagram.  \cite{Brammer+09} report that the extinction-corrected $(U-V)_{\rm rest}$ color performs well in distinguishing dusty star-forming galaxies from red galaxies.
Inspired by \cite{Brammer+09}, \cite{Wang+17} build the following separation criteria to divide galaxies into blue, green, and red galaxy populations:
\begin{mathletters}
\begin{eqnarray} \nonumber
 (U - V )_{\rm rest} - \Delta A_V =   0.126 \log (M_*/M_{\odot}) + 0.58 - 0.286z; \,\, \nonumber  \\
\nonumber
 (U - V )_{\rm rest} - \Delta A_V =   0.126 \log (M_*/M_{\odot})  - 0.24 - 0.136z, \,\, \nonumber
\end{eqnarray}
\end{mathletters}
where $\Delta A_V$ is the extinction correction of rest-frame $U - V$ color, which is equivalent  to $ 0.47 \times A_V$. The correction factor 0.47 is determined for the \cite{Calzetti+00} extinction law. Although the exact  attenuation, $A_V$, suffers from the degeneracy between stellar age and dust attenuation,  this dust-corrected rest-frame $U - V$ color, $(U - V )_{\rm rest} - \Delta A_V$, is able to allow a reliable separation of the galaxies in different star-formation states, which will be demonstrated in the $UVJ$ diagram (see Section 3.2).

We will adopt the same separation criteria as in \cite{Wang+17}. Figure \ref{fig1} shows the extinction-corrected rest-frame colors of all massive galaxies as a function of stellar mass for four redshift bins. The separation lines are shown in orange.
The galaxies above the upper separation line are termed as red galaxy population, and those below the lower separation line are termed as the blue galaxy population.
The region between two separation lines is defined as the GV. 
The above definitions of BC, GV, and RS take into account the redshift dependence of the typical colors of galaxies in those three subpopulations, which is more consistent with our understanding that the high-$z$ galaxies tend to be bluer than low-$z$ ones even for quiescent galaxies.
It can be seen that these selection criteria  perform well in constructing three galaxy populations at $0.5 < z < 2.5$.
As a result, our sample contains 3358 blue galaxies in BC, 2200 green galaxies in GV, and 2686 red galaxies in RS. For each redshift bin, the numbers of galaxies in the three subpopulations are large enough for further statistics.

\section{Physical properties of  massive galaxies}
\subsection{Distributions of Redshift, Stellar Mass, and Dust Attenuation}

\begin{figure*}
\centering
\includegraphics[scale=0.6]{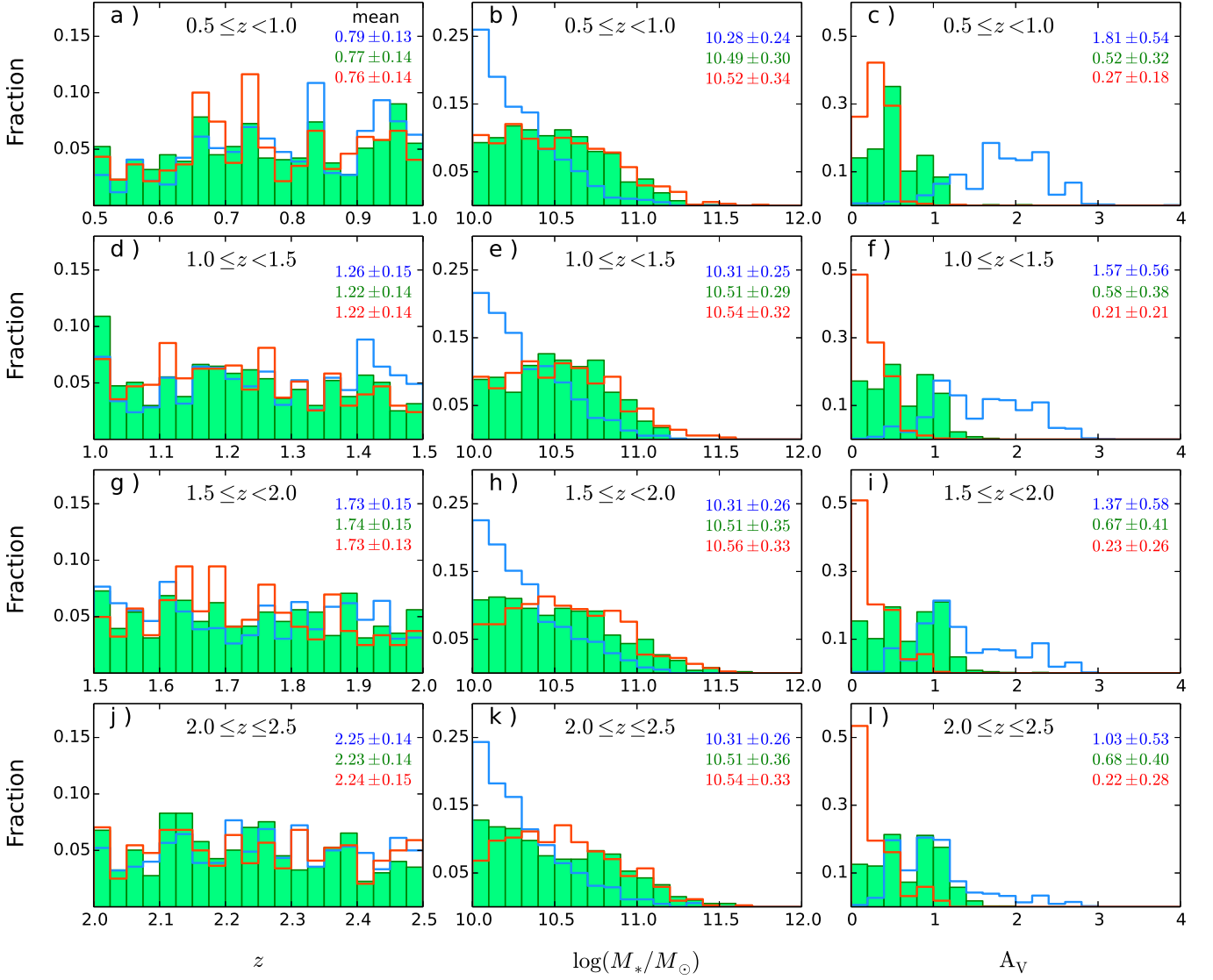}
\caption{Distributions of redshift, stellar mass, and dust-correction factor for red, green, and blue galaxies in various redshift bins, denoted by red, green, and blue solid lines, respectively. Redshift increases from top to bottom.
At the top right of each panel, the mean value and standard deviation are given.}
\label{fig2}
\end{figure*}

The distributions of redshift  ($z$), stellar mass ($\log M_*$), and dust attenuation ($A_V$) for the blue, green, and red galaxies  in each redshift bin are presented in  Figure \ref{fig2}. Mean values and standard deviations are given at the top right of each panel. 
The redshift distributions for three galaxy populations are shown in the left panels of Figure \ref{fig2}. In each redshift bin, the distributions of the three galaxy populations are similar. 

The middle panels show the distributions of stellar mass $\log M_*$.
The difference in stellar mass distribution between red and green galaxies is negligible for whole redshift range. 
However, the stellar mass distributions of the blue galaxies do show a difference compared to those of the green and red galaxies: in all four redshift bins, there is a larger proportion of blue galaxies with relatively lower stellar masses.
Since the probability for a galaxy being quenched is found to increase with stellar mass (\citealt{Peng+10, Brammer+11, Muzzin+13}), it is reasonable that a small proportion of blue galaxies is found at the high-mass end. 

The right panels show the distributions of dust attenuation $A_V$. It is clear that blue galaxies have the largest dust attenuation,  while the least amount of dust is found in red galaxies.
Green galaxies at $0.5<z<2.5$ have an intermediate dust attenuation, which is consistent with the results in the local universe by \cite{Schiminovich+07}, showing that green galaxies are dust attenuated as something in between.
Considering that cool gas is generally associated with dust in star-formation regions, it supports the picture that the green galaxy population is in a transitional state, during which the star-forming galaxies are being quenched into quiescent galaxies.

\begin{figure*}
\centering
\includegraphics[scale=0.72]{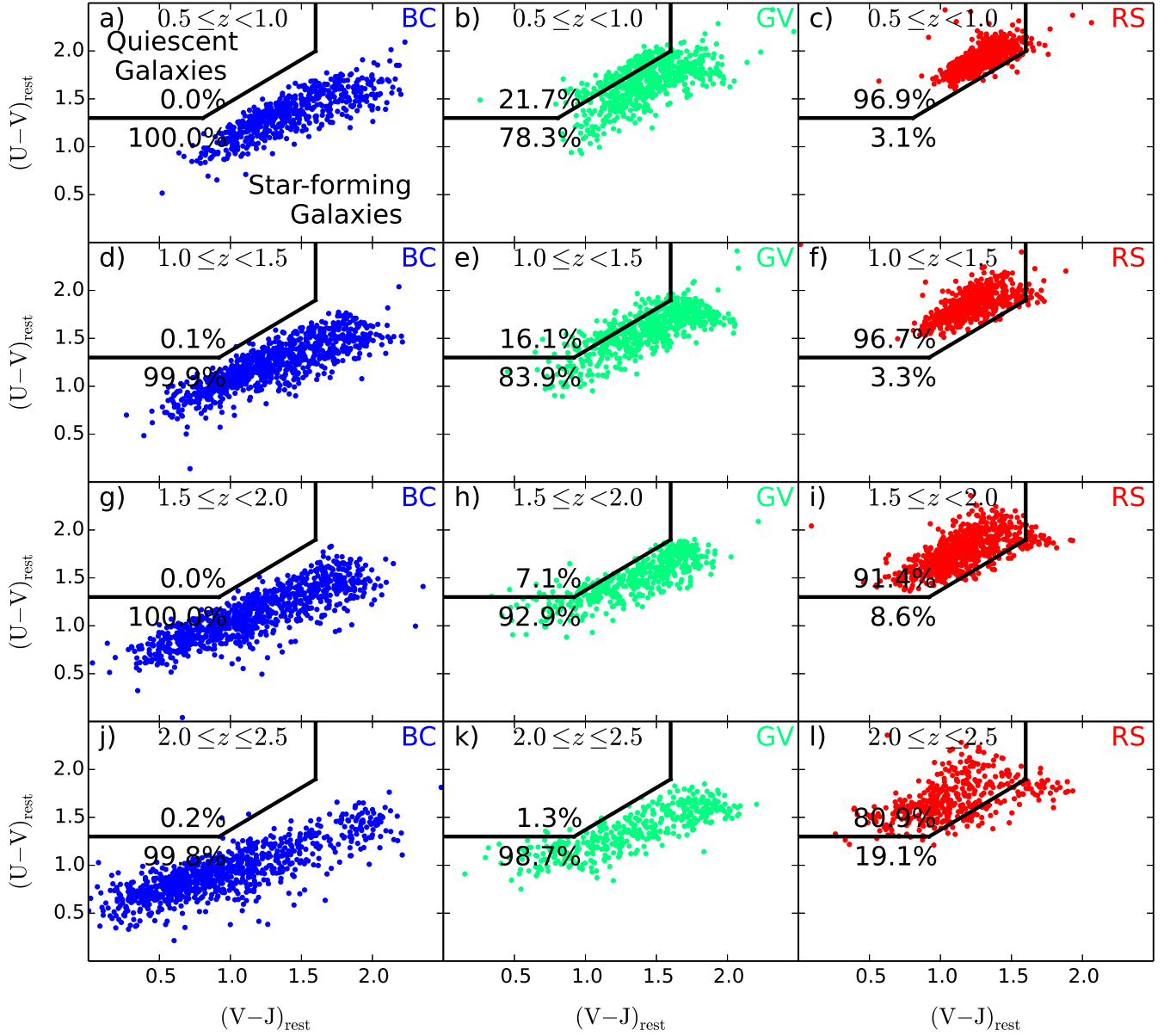}
\caption{Rest-frame $UVJ$ diagram for blue galaxies (left panels), green galaxies (middle panels), and red galaxies (right panels) in four redshift bins, from top to bottom, based on extinction-corrected rest-frame $U-V$ separation. The red , green, and blue circles represent red, green, and blue galaxy populations, respectively. The solid black lines divide massive galaxies into star-forming and quiescent galaxies, following \cite{Williams+09} . The fractions of quiescent and star-forming galaxies are shown near the boundaries.
}
\label{fig3}
\end{figure*}

\subsection{The UVJ Diagram}

The $UVJ$ diagram is a powerful diagnostic of stellar populations in distant galaxies and is widely used to separate galaxies into star-forming  and quiescent galaxies (e.g., \citealt{Wuyts+07, Williams+09, Patel+12, Whitaker+12, Fang+17}).  It is well known that the dusty star-forming galaxies have similar rest-frame $U-V$ colors to the quiescent galaxies.
\cite{Brammer+09} report that the combination of medium-band near-infrared (NIR) filter and  IRAC broadband photometry may trace the slope of the SED redward of the Balmer/4000 \AA\ break (i.e., the rest-frame $V - J$ color), which allows us to separate dusty star-forming and passive galaxies with similar $U-V$ colors.
In order to check the reliability of our samples of red, green, and blue galaxies, we present a rest-frame $U-V$  versus  $V-J $ diagram in Figure \ref{fig4}.
The unobscured star-forming galaxies have both blue $U-V$ and $V-J$ colors, and the  dusty  star-forming galaxies have redder $V-J$ color than quiescent galaxies.
Thus, quiescent galaxies reside in the wedged region in the top left corner, and star-formation galaxies reside in the remaining areas.
Given any aforementioned subsample of red, green, or blue galaxies selected using extinction-corrected rest-frame $U-V$ color$-$mass criteria, they are further divided into quiescent and star-forming galaxies according to the $UVJ$ criteria by \cite{Williams+09}.
The quiescent fraction $f_{\rm Q}$ is defined as $N_{\rm inWedge} /N_{\rm total}$, where $N_{\rm inWedge}$ is the number of galaxies within the $UVJ$ quiescent wedge for a given redshift bin. We have the star-forming fraction $f_{\rm SF}=1 - f_{\rm Q}$.  These two fractions are shown on both sides of the boundary.

It can be found that, in general, our red, green, and blue populations  are in a good agreement with the results of $UVJ$ classification.
Most red galaxies distribute in the quiescent region (top left), and nearly all blue galaxies in the star-forming region.
The green population is located between the red and blue populations in the $UVJ$ diagram. The quiescent fraction of green galaxies increases significantly from a very low value (1.3\%) at $2.0 \leq z \leq 2.5$ to $\sim 20\%$ at $0.5 \leq z<1.0$. This implies that a larger fraction of green galaxies in the local universe may have been quenched to a greater extent.

Assuming that the red galaxies in our sample are the quiescent galaxies, the purity and completeness of quiescent galaxy sample selected with the $UVJ$ diagram can be verified quantitatively. 
The quiescent fraction $f_{\rm Q}$, which we defined above, represents the completeness of $UVJ$-selected quiescent galaxies additionally.
Furthermore, for a given redshift bin, the purity can be defined as the number of our red galaxies within the $UVJ$ quiescent wedge divided by the total number of $UVJ$-selected quiescent galaxies.
Figure \ref{fig4} shows that completeness and purity of the $UVJ$-selected sample of quiescent galaxies vary dramatically with redshift.  For the highest-redshift bin, $2.0 \leq z \leq 2.5$, only a very small percent of green galaxies are misclassified as quiescent; thus the purity of the quiescent galaxy sample is very high ($\sim 99\%$). 
However, about one-fifth of quiescent galaxies at $2.0 \leq z < 2.5$ are excluded by the $UVJ$ classification, which leads to a completeness of $\sim 81\%$.  
For the $UVJ$-defined sample of quiescent galaxies at $0.5 \leq z < 1.0$, a certain fraction of green galaxies distribute in the quiescent region; thus, a very high completeness ($\sim 97\%$) and a purity of $\sim83\%$ are achieved.  
Therefore, the $UVJ$-defined sample of quiescent galaxies in the local universe is probably very complete, but it may include some partially quenched, green galaxies.

\section{Morphology classifications}

By performing the algorithm to the $H$-band images based on convolutional neural networks (see \citealt{Dieleman+15}, for more details),  a morphology catalog of $\sim$50,000 galaxies with $H_{\rm F160W} <$ 24.5 has been presented for the five CANDELS fields by \cite{HCPG+15}. In the current work this catalog of morphological classification will be taken for further analysis.
The machine learning algorithm is trained with the visual classifications in GOODS-S and then applied to the other four fields. To define the morphological class, five parameters (i.e., $f_{\rm spheroid}, f_{\rm disk}, f_{\rm irr}, f_{\rm PS}, f_{\rm Unc}$) for each galaxy are retrieved through analysis of its $H$-band image. These five parameters range from 0 to 1, which represent the probabilities of having a spheroid, having a disk, having some irregularities, being a point source, and being unclassifiable, respectively.
Inspired by \cite{HCPG+15}, massive galaxies in our sample are classified into four morphological classes: (1) spheroid galaxies (bulge dominated), (2) early-type disk galaxies (bulge dominated and having a disk), (3) late-type disk galaxies (disk dominated), and (4) irregular galaxies (including irregular disks and mergers).
In this paper, these  four typical morphological classes are specified as SPH, ETD, LTD, and IRR, for short.
The definition of classification is shown as follows:
\begin{enumerate}
\item Spheroids (SPH): $f_{\rm spheroid} > 2/3$, $f_{\rm disk} < 2/3$, and $f_{\rm irr} < 0.1$ ;
\item Early-type Disks (ETD): $f_{\rm spheroid} > 2/3$, $f_{\rm disk} > 2/3$, and $f_{\rm irr} < 0.1$ ;
\item Late-type  Disks (LTD): $f_{\rm spheroid} < 2/3$, $f_{\rm disk} > 2/3$, and $f_{\rm irr} < 0.1$;
\item Irregulars (IRR): $f_{\rm spheroid} < 2/3$ and $f_{\rm irr} > 0.1$.
\end{enumerate}
\normalsize

\begin{figure*}
\centering
\includegraphics[scale=0.30]{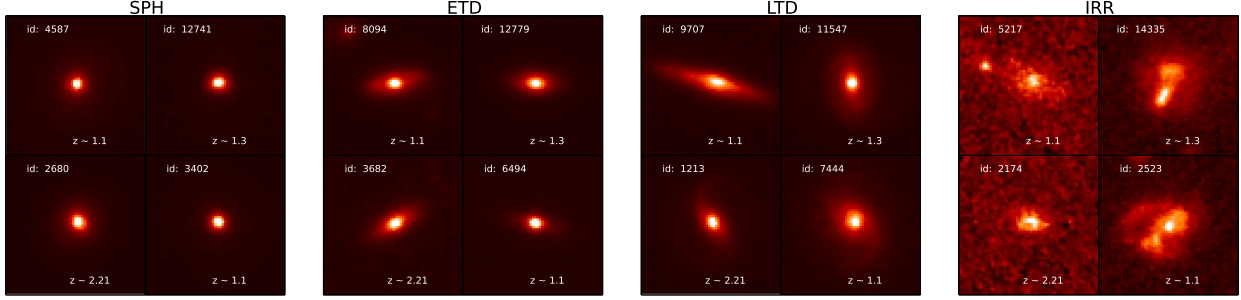}
\caption{Representative image stamps for the galaxies with different morphologies in the GOODS-S field. There are four morphological types, i.e., spheroid (SPH), early-type disk (ETD), late-type disk (LTD), and irregular (IRR), which are presented from the left to right panels. Four stamps is given for each typical morphology, and the stamp size is $4''.0 \times 4''.0$. The ID number and redshift are shown in each image stamp. }
\label{fig6}
\end{figure*}

About 90\% of the massive galaxies in our sample have been classified into these four typical morphologies. The remaining galaxies are classified into the above four typical morphologies (SPH: 38\%; ETD: 15\%; LTD: 13\%; IRR: 34\%) with eyeballing inspect by ourselves.  Roughly half of the remaining galaxies are simply not satisfied by any of the four classifications (e.g., $f_{\rm spheroid} > 2/3\,\, {\rm AND}\,\, f_{\rm irr} > 0.1$). The another half were not measured because they are faint ($H_{\rm F160W} > 24.5$) or located so near to the boundary of coverage. 
Figure \ref{fig6} shows some representative image stamps for the SPH, ETD, LTD, and IRR galaxies in GOODS-S.

\begin{figure*}
\centering
\includegraphics[scale=0.6]{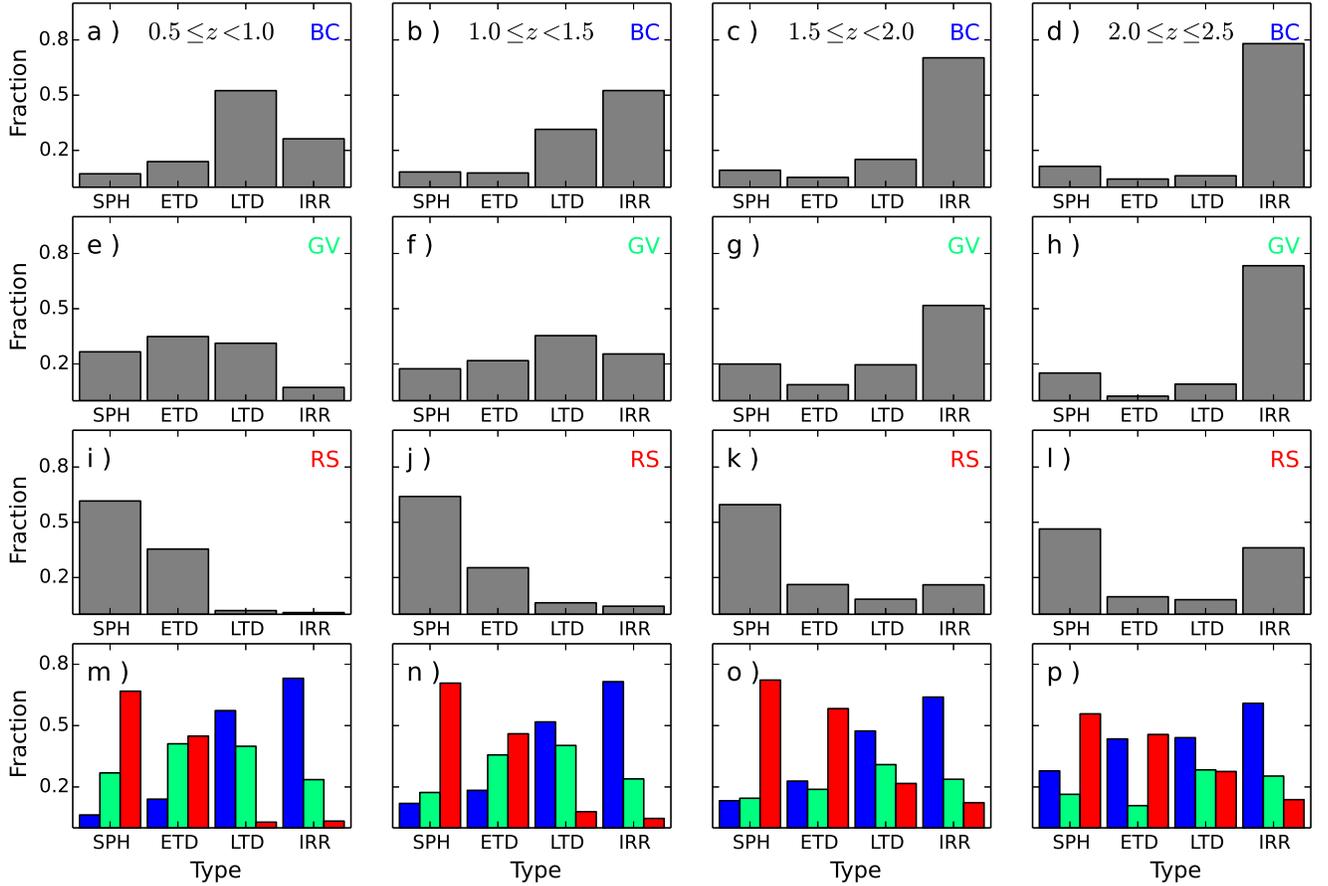}
\caption{Fractions of spheroid  (SPH), early-type disk (ETD), late-type disk (LTD), and irregular (IRR) galaxies in samples of blue, green, and red galaxy populations, respectively.
The first three rows, from top to bottom, present the morphological fractions in four $z$ bins for the BC, GV, and RS galaxies, respectively.
The last row shows the percentages of each galaxy population for the SPH, ETD, LTD, and IRR galaxies in four $z$ bins. The RS, GV, and BC subsamples are denoted with the red, green, and blue colors, respectively.}
\label{fig7}
\end{figure*}

The main structure of distant galaxies (e.g., bulge and disc) can be well quantified by the S\'{e}risc  index ($n$), which is equivalent to the Hubble sequence.  Up to $z  \sim 2.5$,  the blue star-forming galaxies can be well described by exponential disks ($n \sim 1$), while the shape of quiescent galaxies can be better approximated by de Vaucouleurs profiles ($n\sim 4$) (\citealt{Wuyts+11}).
The histograms of morphological classification for our sample are presented in Figure \ref{fig7}, with increasing redshift ranges from left panel to the right.
Panels (a)$-$(l) show the morphological fractions in four redshift bins for the BC, GV, and RS galaxies, respectively.
The correlation between morphology and stellar population is found to be in place since at least $z  \sim 2.5$ by \cite{Wuyts+11}. For the blue galaxies at $0.5 \leq z <1.0$,  more than 50\% are disk dominated (i.e., classified as LTD). For the blue galaxies at higher redshifts, the IRR galaxies come to be predominant. Majority of red galaxies at $0.5\leq z \leq2.5$ are classified as SPH and ETD types, which means that their spheroids are clearly detected. 
Many previous works found that, for the GV galaxies, early-type spirals (Sa$-$Sbc) and the lenticulars (S0s, bulge-dominated disk galaxies) predominate in the local universe (\citealt{Salim+14, Bait+17}) and at intermediate redshift (e.g., $0.4<z<1.2$ in \cite{Mendez+11}). For the green galaxies at $0.5 \leq z < 1.5$ in our sample, more than 50\% are classified as the ETD and LTD types. The presence of bulge and disc structures in green galaxies is also found at $0.4<z<1.2$ by \cite{Mendez+11}.  At higher redshifts, $z \geq 1.5$, the green galaxies are found to be dominated by the IRR galaxies. In general, the morphological types of green galaxy population are intermediate between red and blue populations, which is consistent with the results at $0.2<z<2.0$ by \cite{Ichikawa+17}.  It implies that the transformation from blue to red populations is accompanied by the growth of the  bulge component. Additionally, for all three galaxy populations, an increasing trend in the IRR proportion with the increase of redshift is remarkably shown in panels (a)$-$(l).

Panels (m)$-$(p) in Figure \ref{fig7} exhibit the percentages of each galaxy population for the SPH, ETD, LTD, and IRR galaxies in different redshift bins.
In general, the red galaxy population predominates in the SPH and  ETD types, while the blue galaxy population predominates in the IRR and LTD types.
For each redshift bin, the proportion of blue galaxies increases as we go from the early to late types (i.e., from SPH to IRR), while the proportion of red galaxies exhibits the  opposite trend.
Moreover, over cosmic time (from high $z$ to low $z$), a larger fraction of blue galaxies appear to be the LTD and IRR galaxies, and a larger fraction of red galaxies are found to be the SPH and ETD galaxies with striking spheroids.
The green galaxies at $0.5 \leq z <2$ in various morphologies have a roughly intermediate proportions between the red and blue galaxies. The correlation between morphology and galaxy population is rather significant, which is consistent with the morphology$-$sSFR relation in the local universe (\citealt{Bait+17}).  It points to the scenario from another angle that the buildup of bulges is in progress when blue star-forming galaxies are being quenched to red.

\section{Structure of massive galaxies}

\subsection{Parametric Measurements}
\begin{figure*}
\centering
\includegraphics[scale=0.6]{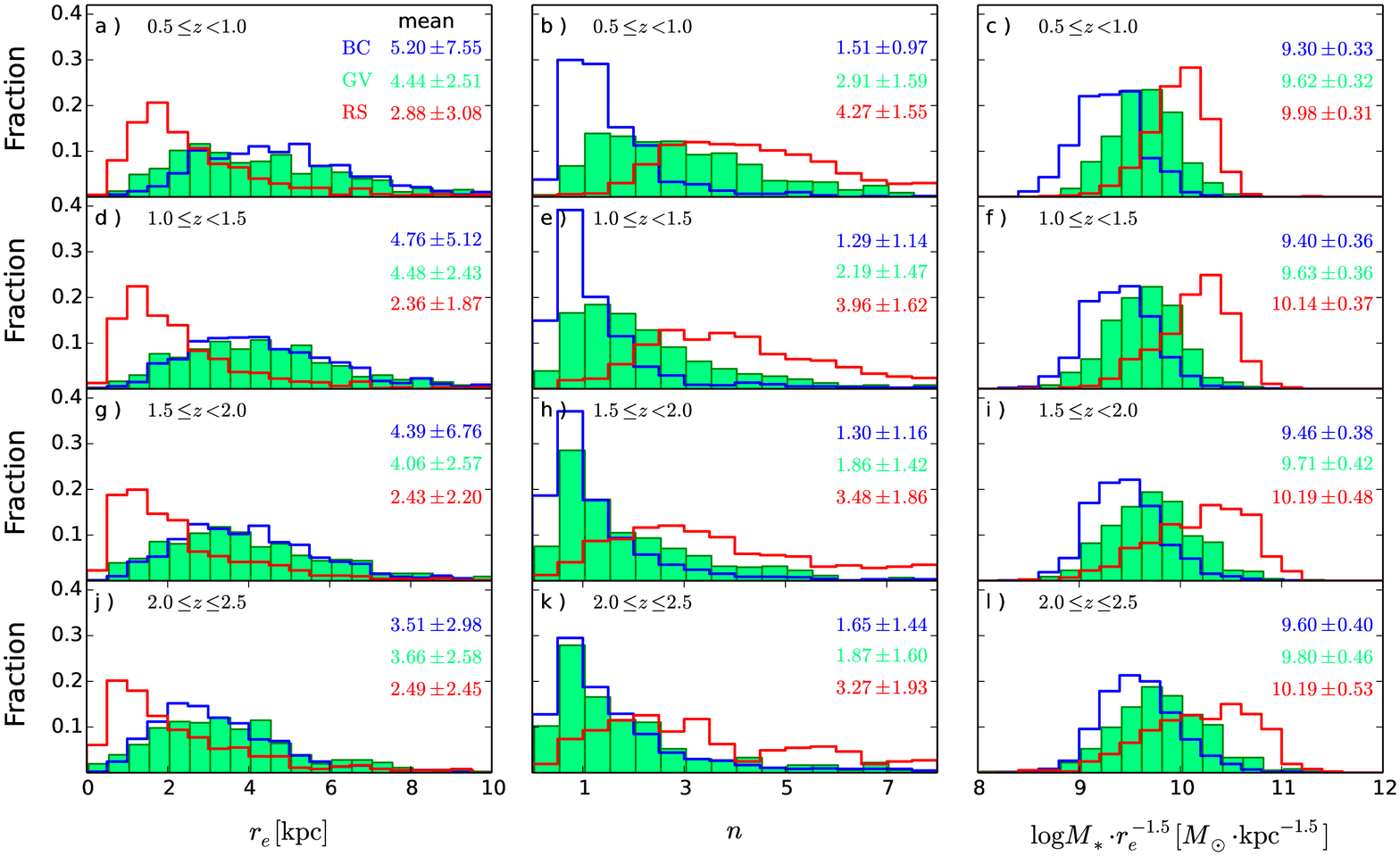}
\caption{Distributions of effective radius, S\'{e}rsic index, and  compactness of blue (blue), green (light green filled), and red (red)  galaxies, with redshift increasing from top to bottom.
The median values of these properties are shown in different colors, respectively.
}
\label{fig8}
\end{figure*}

\begin{figure*}
\centering
\includegraphics[scale=0.5]{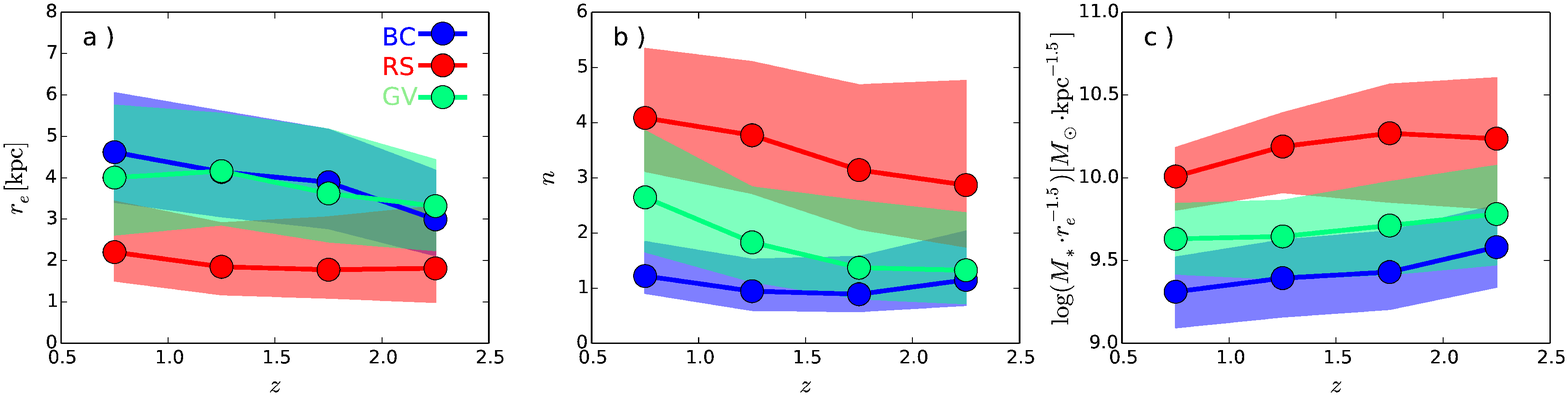}
\caption{The evolution of effective radius, S\'{e}rsic index, compactness of blue (blue), green (light green filled), and (red)  galaxies with redshift.
The circles show the median values in redshift bins, and the shaded regions reflect the 25 to 75 percentiles.
}
\label{fig9}
\end{figure*}

Galaxy structural parameters, such as S\'{e}rsic index  ($n$) and effective radius ($r_e$), have been measured in CANDELS by \cite{vdW+12}. 
Each galaxy is modeled with a single S\'{e}rsic profile in NIR images by GALFIT (\citealt{Peng+02}). Details on measurements were described by \cite{vdW+12}.   
The Rainbow database
\footnote{\url{http://rainbowx.fis.ucm.es/Rainbow\_navigator\_public/}}
has matched the 3D-$HST$ catalog with this catalog of $J$-band and $H$-band.
Considering that the structures of galaxies observed are dependent on the filters, we trace the rest-frame optical morphologies across the redshift range $0.5 < z < 2.5$ by using $J$-band (F125W) imaging for galaxies with $z <1.8$ and $H$-band (F160W) imaging for galaxies with $z > 1.8$. To ensure a credible GALFIT  measurements and maximize the sample size, we choose the galaxies that are constrained to be not bad (quality flag = 0 or 1).  
We also calculate the compactness of a galaxy defined as  $\Sigma_{1.5} = \log{M_*/r_e^{1.5}}$, as in \cite{Barro+13}.

Figure \ref{fig8} shows that the distributions of $r_e$, $n$ and $\Sigma_{1.5}$  of blue (blue), green (light green filled), and red (red)  galaxies, with redshift increasing from top to bottom.
The evolution of these three parameters with redshift is shown in Figure \ref{fig9}. 

\begin{figure*}
\centering
\includegraphics[scale=0.56]{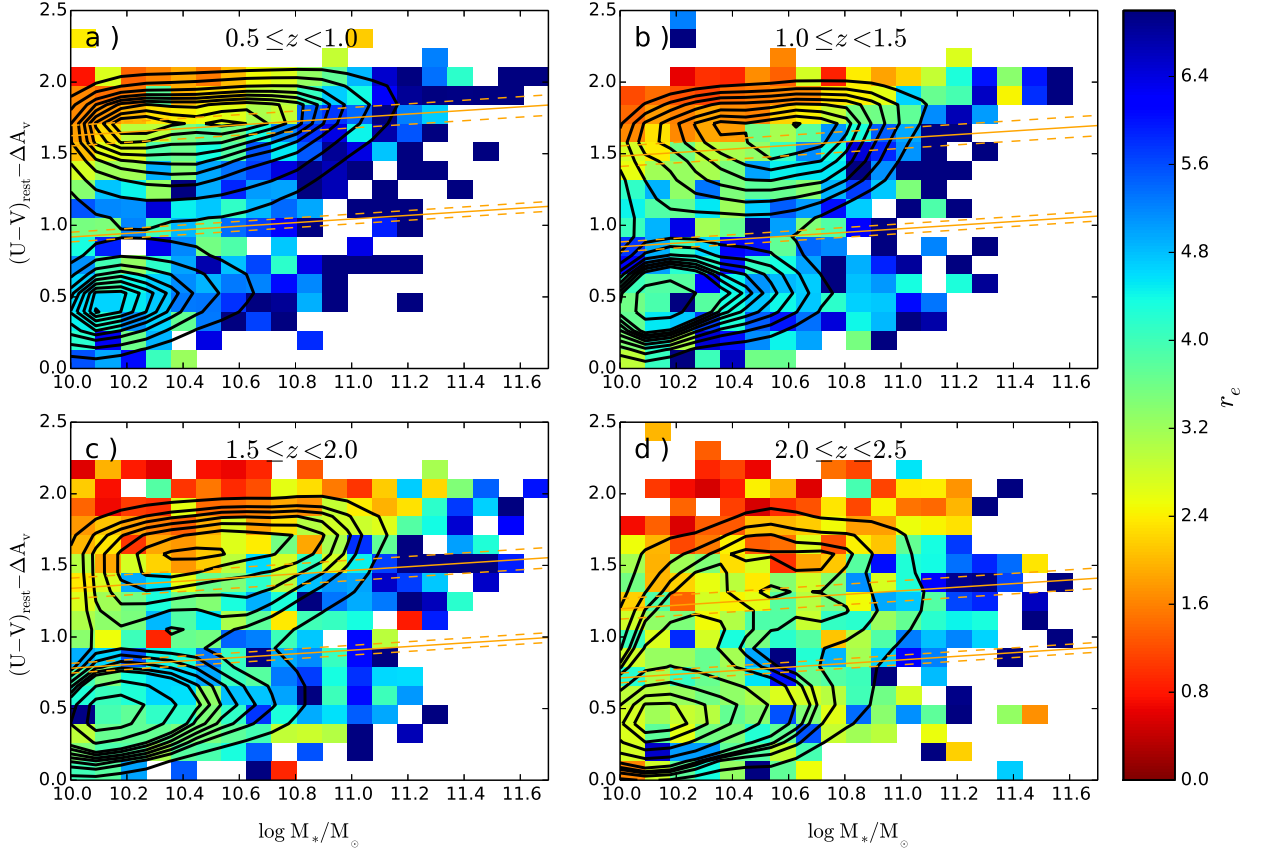}
\caption{Extinction-corrected rest-frame $U-V$ color as a function of stellar mass in four redshift bins.
As in Figure \ref{fig1}, the contours represent the relative density in the color$-$mass space. The criteria for each redshift bin are shown as orange solid lines. Color blocks represent the mean $r_e$. 
}
\label{fig1_re}
\end{figure*}

We find that the red galaxy population has the lowest $r_e$, but the green and blue galaxy populations are hard to differentiate. However, \cite{Pandya+17} find that green galaxies have an intermediate $r_e$ between blue and red galaxies by studying mass-match subsamples of three galaxy populations. In their work, each green galaxy matches three red galaxies and three blue galaxies in the same redshift bin whose stellar masses are within a factor of two. 
Noticing that green galaxies ($\sim 10^{10.5} M_{\odot}$) are averagely more massive than blue galaxies ($\sim 10^{10.3} M_{\odot}$) in our sample, it is possible that the conflict is caused by the degeneracy between size and mass. 
\cite{vdW+14} give the size$-$mass relations from $z \sim 3$ to the present epoch but only for star forming galaxies and passive galaxies, respectively. 
To investigate the size dependence of stellar mass, we show the $r_e$ distribution in the color$-$mass diagram in Figure \ref{fig1_re}. In each panel, galaxies with higher $\log M_*$ and bluer $U - V$ color tend to be larger in size. A strong correlation between galaxy structures and star formation activities is found in the SFR$-M_*$ plane (\citealt{Wuyts+11,Brennan+17}), which points to a similar trend in models and observations. 

\begin{figure*}
\centering
\includegraphics[scale=0.52]{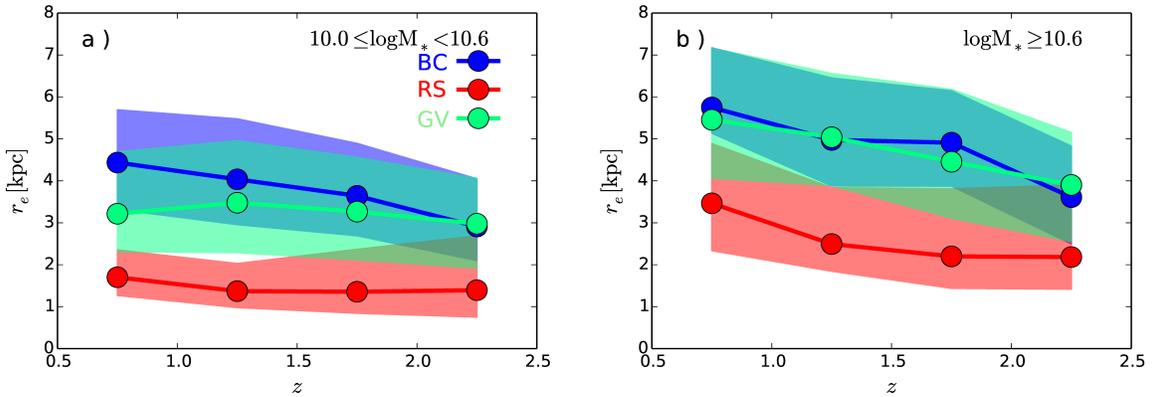}
\caption{Effective radius for our three galaxy populations with stellar mass below (left panel) and above (right panel) $\log(M_*/M_{\odot}) = 10.6$ as a function of redshift of our sample. 
}
\label{fig10_2bin}
\end{figure*}

As shown in Figure \ref{fig1_re}, the $r_e$ difference among three galaxy populations is more prominent at the low-mass end.  
To show this evidence more clearly, we simply split our sample into the low-mass ($10.0 \leq \log M_* < 10.6$) and high-mass ($\log M_* \geq 10.6$) subsamples and show their size$-$ redshift relations in Figure \ref{fig10_2bin}. 
Among the most massive galaxies with $M_* \geq 10^{10.6} M_{\odot}$, no significant difference of size ($r_e$) is found between green and blue galaxies.
However, for the less massive galaxies, green and blue galaxies have a similar median $r_e$ at $z > 2.0 $, and tend to be differentiable over cosmic time since $z \sim 2$. This result seems to be insensitive to the divisor for two mass bins because the conclusion does not change when applying a $\pm$ 0.1 dex shift on the mass divisor.


There is a striking gradient of $n$ and $\Sigma_{1.5}$ from red to green to blue galaxies. Red galaxies are more compact and have higher S\'{e}rsic indices, which means that they are bulge dominated.
In contrary, blue galaxies are less compact and have lower S\'{e}rsic indices, which means that they are disk dominated.
Green galaxies are always intermediate in $n$ and $\Sigma_{1.5}$ distributions between red and blue galaxies, which is consistent with the results  in \cite{Pandya+17}.
For red and green galaxies, their medians of S\'{e}rsic index $n$ increase from high to low redshift, but blue galaxies seem to have a constant median, $n \sim 1$, suggesting that the transitional phase is accompanied by the buildup of the bulge component.

\subsection{Nonparametric Measurements}
\begin{figure*}
\centering
\includegraphics[scale=0.6]{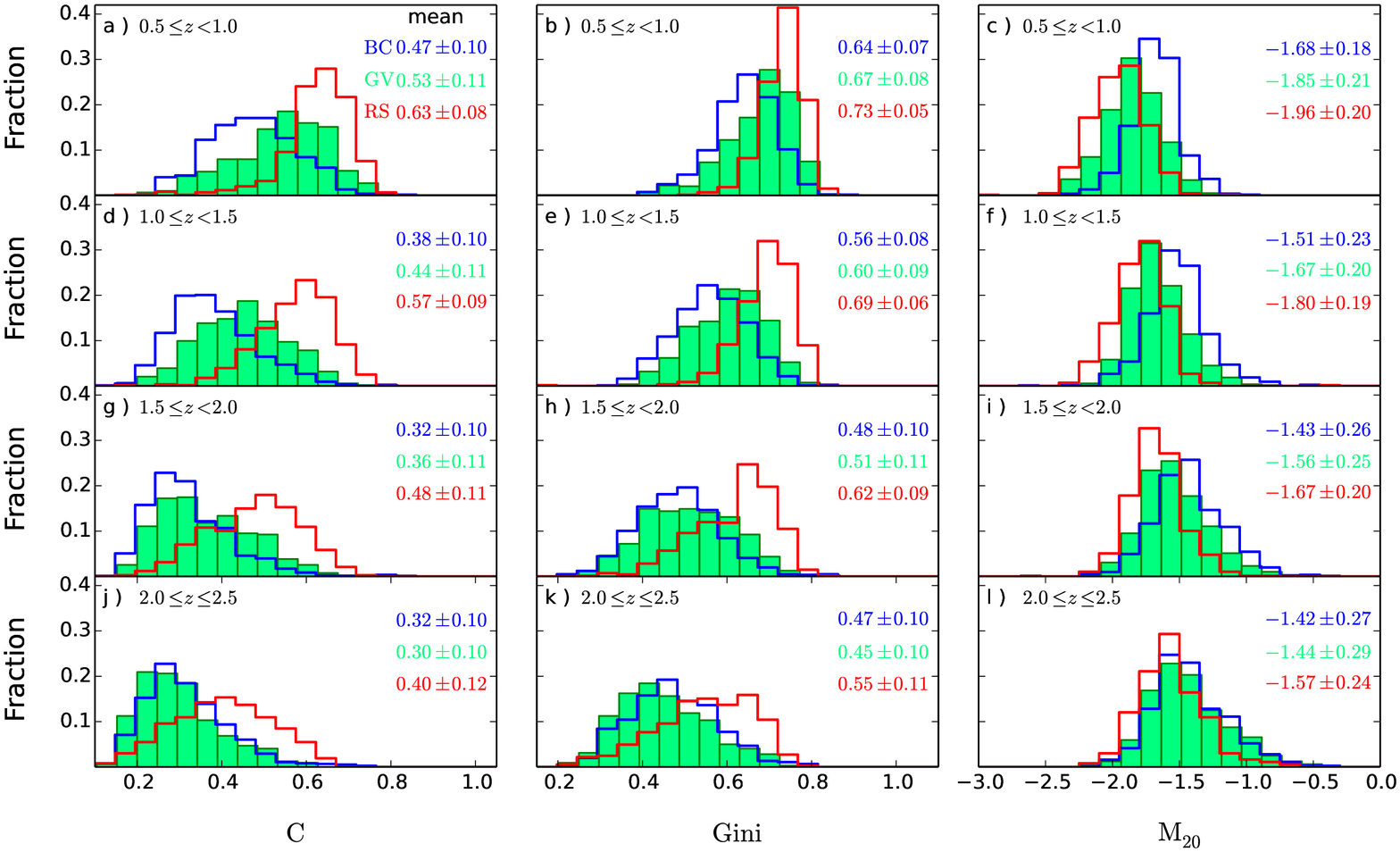}
\caption{Distributions of $C$, $G$, and $\rm M_{20}$ of blue (blue), green (light green filled), and red (red) galaxies, with redshift increasing from top to bottom.
The median values of these properties are shown in the corresponding colors, respectively.
}
\label{fig11}
\end{figure*}

We also performed our own nonparametric structural measurements on the NIR images using the Morpheus software, developed by \cite{Abraham+07}. 
The parameters, including concentration ($C$), Gini coefficient ($G$), and second-order moment of light ($M_{20}$), are introduced to quantify the structures of galaxies. 

Concentration index ($C$) is a parameter that describes the concentration of the surface brightness distribution in a galaxy.
Following \cite{Abraham+94}, the definition of $C$ is the ratio between two integral fluxes within the inner isophotal radius, $0.3R$, and within the outer isophotal radius, $R$, 
\begin{equation}
C = \log (\frac{F_{0.3R}}{F_{R}}),
\end{equation}
where $R$ is the radius for an enclosed area by the galaxy isophote at the 2$\sigma$ level above the sky background. 
Concentration index ($C$) decreases with Hubble sequence from the ellipticals to the irregulars.

The Gini coefficient ($G$) is a statistical tool to quantify the unequal light distribution, which is defined as (\citealt{Lotz+04})
\begin{equation}
    G = \frac{ \sum_{i}^{n}(2i - n- 1)\left| F_i \right|}{\left|\bar{F} \right|n(n-1)},
\end{equation}
where $F_i$ is the pixel flux value sorted in ascending order, $n$ is the total number of pixels uniquely assigned to a galaxy during object detection, and $\bar{F}$ is a mean flux for all the pixels.
It can be seen that the Gini coefficient ($G$) is relative to the concentration index ($C$). For a galaxy with a high concentration index ($C$), the galaxy light is concentrated at the center, and thus the Gini coefficient  ($G$) is likely to be high. Otherwise, the galaxies with high $G$ values may not have a high $C$, because the pixels with higher flux may distribute in the outer region, rather than in the central region.
Compared with the concentration index ($C$), the Gini coefficient ($G$) can be applied to the galaxies with any shapes and at any distances, because a specified central point is not necessary in calculating $G$. Gini coefficient measurements are advantageous for the galaxies at high redshift, where most galaxies are irregular without distinct centers.

The second-order moment of the 20\% brightest pixels ($M_{20}$) is a quantity that facilitates to reveal the presence of substructures, such as bars, spiral arms, and bright cores. It is defined as
\begin{equation}
  M_{20} = \log(\frac{\sum_{i} M_i}{M_{tot}}) , {\rm with} \sum_i F_i < 0.2F_{\rm tot},
\end{equation}
where $M_i = F_i[(x_i - x_c)^2+(y_i - y_c)^2]$ and $M_{\rm tot} = \sum_{i=1}^{N}{M_i} $ for the fluxes of the brightest 20\% of light in a galaxy.
Late-type galaxies have a typical $M_{20}$ of $\sim$ -1.5, and the typical  $M_{20}$ for early-type galaxies is about -2 (\citealt{Lotz+04}).

\begin{figure*}
\centering
\includegraphics[scale=0.5]{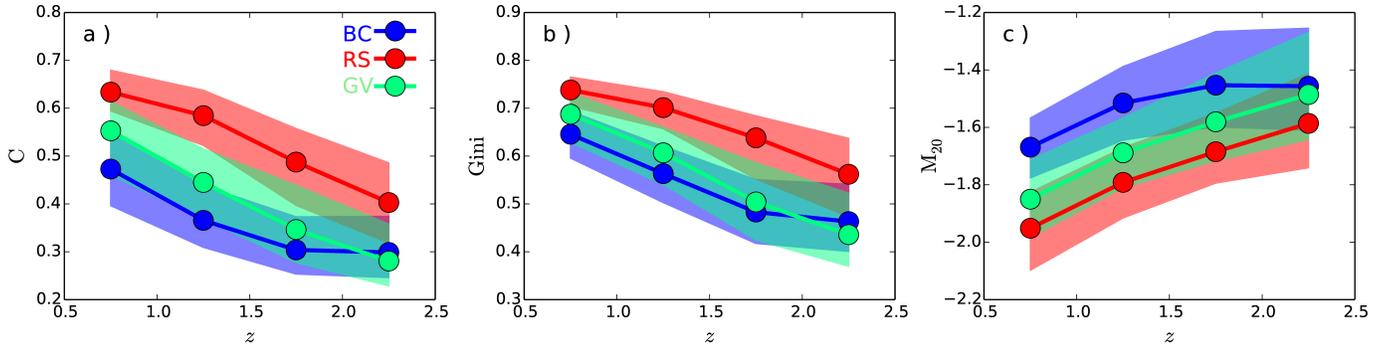}
\caption{Redshift evolutions of $C$, $G$, and $\rm M_{20}$ of blue, green, and red galaxies with redshift. The filled circles denote the median values in various redshift bins, and the shaded regions cover the 25 to 75 percentiles.
}
\label{fig12}
\end{figure*}

Figure \ref{fig11} shows the distributions of $C$, $G$, and $M_{20}$ for three galaxy populations, in an increasing order of redshift from top to bottom. Basically, red galaxies are found to have large concentration indices and Gini coefficients, while blue galaxies have small values of $C$ and $G$. Blue galaxies have the largest $M_{20}$ values for all redshift bins. Green galaxies exhibits exactly intermediate distributions of these three nonparametric measurements. 
The cosmic evolution of $C$, $G$, and  $M_{20}$ for three galaxy populations is given in Figure \ref{fig12}. 
The three galaxy subpopulations show dramatic morphological evolution at $z<2$ in terms of these three nonparametric properties.
It is found that the $C$ and $G$ values of the galaxies at $z<2$ considerably increase over cosmic time, and the $M_{20}$ tends to decrease. It points to an  evolutionary trend that high-$z$ galaxies are more concentrated and regular in shape and their substructures are less prominent.  
This result is consistent with our interpretation in section 4 that bulge growth is accompanied by quenching of star formation.
To understand the morphology evolution for three galaxy populations, both bulge growth and star formation quenching should be taken into account.

Moreover, it should be noticed that the blue and green galaxies at $2.0< z<2.5$ have almost the same distributions of nonparametric measurements. Green and blue galaxies have similar  S\'ersic index distribution at $2.0<z<2.5$ (see the middle panel of Fig. \ref{fig9}). The differences in the $n$ distributions between green and blue galaxies begin to be amplified gradually from $z \sim 2 $ to present day. This implies that, at the early stage of star formation quenching (i.e., $z>2$), the morphological change is not significant. 

\section{AGN fraction}

In order to understand the effect of AGN feedback on star formation activities, we estimate the AGN fractions for three galaxy populations in Chandra Deep Field-North (CDF-N) and Chandra Deep Field-South (CDF-S), covering the two fields of GOODS-N and GOODS-S in our sample. 
The other three fields do not have as deep X-ray data and catalogs over a large enough area on the sky as GOODS-N and GOODS-S. 
\cite{Xue+16} present an improved point-source catalog for the $\sim$ 2 Ms exposure in CDF-N, and \cite{Luo+17} contribute the X-ray source catalog for the $\sim$ 7 Ms exposure in CDF-S. In each field, they provide a main catalog of X-ray sources which have source detections of high significance and a supplementary catalog of X-ray sources with relatively low significances that are identified to be bright in NIR bands. 
The some of these sources that are classified as AGNs just need to satisfy one of the five following criteria: 
(1)$L_{\rm 0.5-7 keV} \geqslant  3 \times 10^{42} \rm{erg s^{-1}}$, ensuring that the X-ray emission is from central AGNs, not from $\rm H_{II}$ regions; 
(2) the effective photon index $\Gamma \leqslant 1.0$, as one signature of moderately to highly obscured AGNs;
(3) the X-ray-to-optical flux ratio $\log( f_X / f_R)> -1$
, where $f_X = f_{\rm 0.5-7 keV}$, $f_{\rm 0.5-2 keV}$, or $ f_{2-7 keV}$, and $f_R$ is the observed-frame $R$-band flux, which is a useful AGN/galaxy discriminator (\citealt{Bauer+04, Xue+10});
(4) spectroscopically classified as AGNs by broad emission lines and/or high-excitation emission lines;
and 
(5) $L_{\rm 0.5-7 keV}/L_{\rm 1.4GHz} \geq 2.4 \times 10^{18} $, where
$L_{\rm 1.4GHz}$ is the rest-frame 1.4 GHz monochromatic luminosity in units of $\rm W\ Hz^{-1}$, indicating an excess X-ray emission over the level expected from starburst galaxies. 
The above criteria were described in detail in \cite{Xue+11}. Although low-luminosity and/or highly obscured AGNs may still be missed, a relative complete AGN sample can be constructed through the criteria.

A matching radius of 1{\arcsec}.5 is adopted to identify the host galaxies of AGNs, and 192 AGNs in GOODS-N and 186 AGNs in GOODS-S have been unambiguously identified. 
The AGN fractions ($f_{\rm AGN}$) of three galaxy populations are simply defined as $f_{\rm AGN} = N_{\rm AGN}/N_{\rm tot} $, where $N_{\rm AGN}$ and $N_{\rm tot}$ are numbers of AGN host galaxies and the massive galaxies in a well-defined sample. The statistical uncertainty of the AGN fraction can be computed by 
$\sigma_f = [f_{\rm AGN}(1-f_{\rm AGN})/N_{\rm tot} ]^{1/2}$, assuming binomial statistics. 
Considering that the X-ray imaging depths are different in two fields, we separately calculate the  AGN fractions for three galaxy populations  in GOODS-N and GOODS-S.  
We will look at the X-ray AGN fraction integrated over our entire redshift range ($0.5<z<2.5$), as well as in two smaller redshift bins.
Since that blue and green galaxies start to show different morphologies (e.g., S\'ersic index $n$) and structures (e.g., $C$, $G$, and $M_{20}$) at $z = 2$,  we divide our sample into high-$z$ ($2.0< z<2.5$) and low-$z$ ($0.5<z<2.0$) subsamples in order to observe the redshift evolution of the AGN fraction.  

\setlength{\tabcolsep}{2pt}
\setcounter{table}{0}
\begin{table}[h!]
\renewcommand{\thetable}{\arabic{table}}
\centering
\caption{AGN fractions in GOODS-N and GOODS-S} \label{tab1}
\begin{tabular}{ccccc}
\tablewidth{0pt}
\hline
\hline
Field  & $N_{\rm tot}$ &BC & GV & RS \\
\hline
$0.5<z<2.5$& &    &    &    \\
GOODS-N   &1457&$11.5\%  \pm 1.2\%$&$20.9\% \pm 2.2\%$&$  9.5\% \pm 1.4\%$\\
GOODS-S   &1368&$  9.8\%  \pm 1.3\%$&$19.6\% \pm 2.1\%$&$13.2\% \pm 1.5\%$\\
\hline
   $0.5 < z < 2.0$ & &       &        &       \\
GOODS-N   &1132&$13.3\% \pm 1.6\%$&$21.7\% \pm 2.4\%$&$   7.9\%\pm 1.4\%$\\
GOODS-S   &1098&$10.5\% \pm 1.6\%$&$20.5\% \pm 2.4\%$&$12.3\%\pm 1.6\% $\\
\hline
   $2.0 \leq z < 2.5$ & &      &        &       \\
GOODS-N  &325&$7.1\% \pm 1.8\%$&$15.7\% \pm 5.1\%$&$16.9\% \pm 4.3\% $\\
GOODS-S  &270&$7.6\% \pm 2.3\%$&$15.7\% \pm 4.4\%$&$18.8\% \pm 4.7\% $\\
\hline
\end{tabular}
\label{tab1}
\end{table}

It has been reported that the AGN detection rate is higher in GV galaxies, regardless of whether the AGN is selected  by X-ray observation (\citealt{Nandra+07}) or optical line-ratio diagnostics (\citealt{Salim+07}). The AGN fraction and its uncertainty for each subsample are given in Table \ref{tab1}.
It is shown that the AGN fraction for the green galaxy population at $0.5<z<2.5$ is $\sim 20\%$, which is remarkably larger than those for the red and blue populations.
Green galaxies at high redshifts are found to have a greater probability of hosting AGN activity. For the AGN host galaxies in the local universe, \cite{Schawinski+10} find that they  are preferentially located in GV regions in the color$-$mass diagram. 

For the low-$z$ subsamples ($0.5<z<2$),  red and blue galaxies have similar AGN fractions,  significantly lower than that of green galaxies at $z<2$. 
AGN feedback may heat the interstellar media (ISM) and thus restrain the subsequent star formation in the AGN host galaxy. 
If GV galaxies are a transitional population where star formation is being partially quenched, a larger AGN fraction in the green galaxy population seems to favor the scenario that AGN feedback plays an important role in suppressing the global star formation.    
It is interesting to see that red galaxies at $z>2$ are found to have a higher AGN fraction, $\sim$ 16\%, comparable to that of green galaxies, despite of the small sizes of high-$z$ subsamples.  
\cite{Wang+17} also find a high proportion of AGNs in the red galaxy population at high redshift. 
Although it is hard to conclude that AGN feedback directly results in the quenching of global star formation,  
it still implies that  AGN feedback may play important roles not only in transforming star-forming galaxies to quiescent galaxies but also in keeping the quiescence of galaxies at high redshift.

\section{Environmental Effect}

\begin{figure*}
\centering
\includegraphics[scale=0.6]{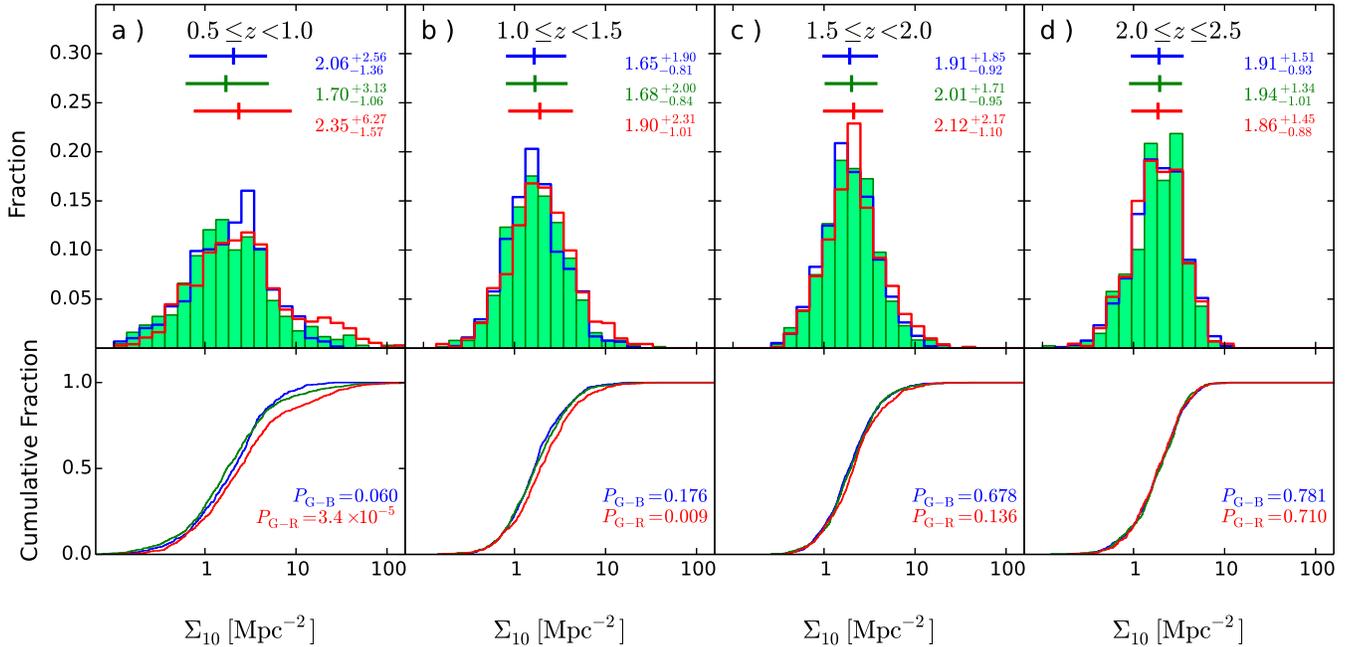}
\caption{Distributions of local surface density ($\Sigma_{10}$) for blue (blue line), green (light green filled), and red galaxies (red line) in four redshift bins in the top panels.
At the top right of each panel, median $\Sigma_{10}$ values are given. The error bars denote the median of the offsets from the median value.
The bottom panels show the cumulative fraction distributions for blue (blue line), green (light green line), and red galaxies (red line). 
The probabilities of K-S tests between red and green galaxies are shown in red, and the results between blue and green galaxies are shown in blue. 
}
\label{fig13}
\end{figure*}

We define the local surface density by the nearest 10 neighboring galaxies, $\Sigma_{10}$, as an indicator of the local environment, which was firstly proposed by \cite{Dressler+80}. 
For each galaxy, it can be calculated by $\Sigma_{10} = 11/(\pi D^2_{\rm p,10})$, where $D_{\rm p,10}$ (in Mpc) is the  projected distance to the 10th-nearest neighbor within a redshift slice.
The redshift range of each slice depends on the redshift precision.
Due to different precisions of photometric redshift in five fields, we adopt a mean $z_{\rm phot}$ uncertainty, $\sigma_{\rm z} \sim 0.018$, for all massive galaxies in our sample. 
All the massive ($\log M_* > 10$) galaxies within a redshift slice, $|\Delta z| < \sigma_{\rm z}(1+z)$, will be taken into account when deriving the surface density.  
It should be noticed that the number of nearest neighboring galaxies ($N$) may directly affect the uncertainties in the local surface density estimate. Basically, the third-nearest neighbour is applied when the redshift estimate is very precise and reliable (e.g. spectroscopic redshift in most cases). For the photometric redshift estimate with lower precision, the local density estimate will be more uncertain when taking a small $N$ value.  To ensure the reliability of the local density estimate, following many other works (e.g., \citealt{Capak+07},\citealt{Pan+13}), In our work we conservatively adopt $N=10$ to reduce the uncertainty of line-of-sight distance on average. When we take $N = 9, 11$, no significant difference is found in density distributions between the three subpopulations. 

Distributions of projected surface densities for blue, green, and red galaxies in four $z$ bins are shown in the top panels of Figure \ref{fig13}. 
To compare the environment of the green galaxy population with those of the red and blue populations, we perform the Kolmogorov$-$Smirnov (K-S) tests to see whether green galaxies have different $\Sigma_{10}$ distributions relative to the red and blue galaxies. 
The bottom panels of Figure \ref{fig13} present the cumulative distribution functions for the samples of three galaxy populations in various $z$ bins. 
The quantity $P$ is defined to give the probability that two samples are drawn from the same underlying parent distribution.
We commonly adopt a critical value of $P = 0.05 (=5\%)$ as the upper limit to verify that two samples have different $\Sigma_{10}$ distributions at $\geq 2\sigma$ significance. The probabilities for the K-S tests between green and blue/red galaxy samples are also shown in the bottom panels. 


Figure \ref{fig13} presents some clues on cosmic evolution of environmental diversities among three galaxy populations. The results of K-S tests show that red, green, and blue galaxies at high redshfits ($z>1.5$) have similar $\Sigma_{10}$ distributions. 
Since $z \sim 1.5$, the red galaxy population exhibits a $\Sigma_{10}$ distribution that differ from the green and blue galaxy populations. Red galaxies appear to prefer a dense environment, whereas green and blue galaxies are preferentially found in lower-density environments. This is consistent with the early findings by \cite{BO78} that a certain fraction of red and passive galaxies are located in high-density environments (e.g., galaxy groups/clusters) in the local universe, and a larger fraction of blue galaxies appear in these high-density environments at higher redshifts. 
In panel (a) of Figure \ref{fig13}, it is clear that the red galaxy population at $0.5<z<1.0$ comes to have a peak at the high density end. This seems to be in agreement with the scenario that galaxy clusters form at $z \sim 1.0$, and the majority of member galaxies are red, quiescent massive galaxies, particularly in the core region of a cluster.

The K-S tests between blue and green galaxy populations show that these two populations have similar $\Sigma_{10}$ distributions at $z>0.5$ (see $P_{G-B} > 0.05$ for all redshift bins), which is consistent with the results in \cite{Pan+13}. This suggests that, for massive galaxies with $z>0.5$, the environmental effect on star formation quenching is not so important.
Moreover, the $P$ values between blue and green galaxies show a decreasing trend from $z = 2.5$ to 0.5, which means that it is possible that a significant diversity of environment might emerge between the green and blue populations at $z<0.5$. This implication is consistent with the investigation by \cite{Peng+10} who concluded that environment quenching plays a leading role at $z < 0.5$. 
Last but not least, when we probe the role of environment with the CANDELS data, some intrinsic limitations might exist given the fact that each CANDELS field covers only $\lesssim 0.05\ {\rm deg}^2$ on the sky, which is not sufficient to probe significant overdensities (like clusters) at all redshifts (especially lower redshifts).
 
\section{Discussion}

\subsection{Bulge Growth}
It has been proved that red galaxies have bulge components and bulge-dominated disk galaxies predominate in the GV at low and intermediate redshifts (e.g., \citealt{Mendez+11, Lackner+12, Pan+13, Salim+14, Bait+17}).
The high-quality  $HST$/WFC3  imaging data in five CANDELS fields allow us to explore the relationship between between galaxy structural and star formation properties from $z$=2.5 to $z$=0.5, spanning $\sim 6$ Gyr of cosmic time. 

In many previous studies on star formation quenching at high redshifts,  the galaxies are commonly divided into star-forming galaxies and quiescent ones (e.g., \citealt{Brammer+11, Muzzin+13, vdW+14}). 
Given that most quiescent galaxies are bulge dominated, \cite{Bell+12} argue that a prominent bulge is necessary for star-formation quenching over the past $\sim$ 10 Gyr (since $z \sim 2.2$). 
\cite{Lang+14} find that the bulge mass is a better observable parameter for predicting whether a galaxy is star-forming or quiescent than total stellar mass, spanning the redshift range $0.5 < z < 2.5$. 

Recently, \cite{Ichikawa+17} analyse the morphology of recently quenched galaxies in the Cosmic Evolution Survey (COSMOS) UltraVISTA field at $0.2< z < 2.0$ and find that these galaxies may represent a short transitional phase of evolution, accompanied by buildup of the bulge component.
Although not all galaxies possessing bulges lack star formation, it is quite possible that the internal bulge component is associated with star-formation quenching.
 
Our structural analysis is based on the statistics of three different standpoints: morphological classifications, parametric measurements (e.g., $r_e$,  $n$, and $\Sigma_{1.5}$), and nonparametric measurements (e.g., $C$, $G$, and $M_{20}$).
It shows a clear trend of morphological evolution such that the buildup of the bulge component and the quenching of star formation over time are correlated. 
This trend is supported by the statistics of parametric and nonparametric measurements.
Red galaxies are more compact and have higher $n$, which means that those red galaxies are bulge dominated.
On the other hand, blue galaxies is less compact and have lower $n$, suggesting a disk-dominated nature.  
Green galaxies are commonly found to have an intermediate  $n$-distribution between red and blue galaxies, which was also found by \cite{Pandya+17}. 
From blue to red galaxy populations, light profiles come to be more concentrated and less clumpy. 
We confirm that the morphologies of green galaxies at $z<2$ are intermediate between those of blue and red galaxies. 
The structural changes, as one of the main effects of star formation quenching, suggest that galaxy structure evolves together with quenching progress.
While star-forming galaxies are gradually quenched into quiescent and dead, the buildup of the bulge component is ongoing. One possible scenario is that gas-rich compaction, which consumes gas quickly in the center, leads to the inside-out quenching and the subsequent buildup of a dense stellar core (\citealt{Tacchella+16}). 

\subsection{Downsizing of the Green Valley Evolution}
\begin{figure}
\centering
\includegraphics[scale=0.55]{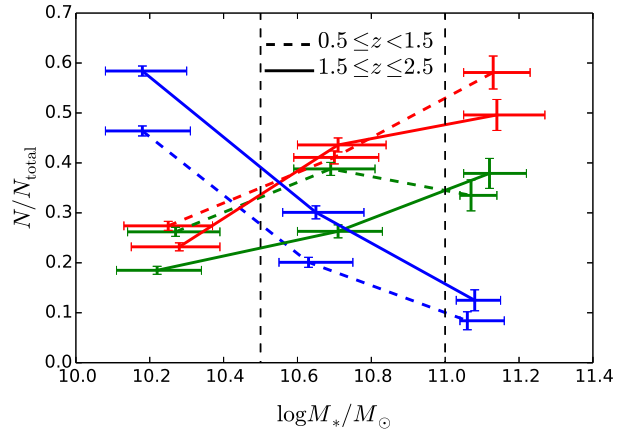}
\caption{Fractions of  red, green, and blue galaxy populations within different stellar mass and redshift ranges. The stellar masses of galaxies, $\log(M_*/M_{\odot})$, are divided into three bins (i.e., [10.0,10.5), [10.5,11.0), and [11.0,11.5]). Two redshift ranges are defined (i.e., $0.5\leq z < 1.5$ and $1.5 \leq z \leq 2.5$). The mass medians in subsamples are shown, and their error bars represent the medians of offsets from the median value.
}
\label{fig3}
\end{figure}

Several mechanisms have been proposed to suppress star formation in massive galaxies, including strangulation (e.g., \citealt{Peng+10}) and AGN feedback (e.g., \citealt{Ciotti+07}). 
The percentages of three galaxy populations within specified stellar mass ranges  may shed light on understanding the degree of star formation quenching as a function of stellar mass.
On the basis of stellar masses, the galaxies are divided into three subsamples with a mass interval of 0.5 dex (i.e., $\log M_*/M_{\odot} \in$ [10.0,10.5),  [10.5,11.0), and [11.0,11.5]). Two redshift bins are defined (i.e., $z \in [0.5, 1.5)$ and [1.5, 2.5]) for studying redshift evolution.
Figure \ref{fig3} shows the fractions of three populations within different mass and redshift ranges. 
In the mass range of  $10 \leq \log (M_*/M_{\odot}) \leq 10.5$, blue galaxies predominate, while red galaxies are predominant in the high-mass range,  $11 \leq \log (M_*/M_{\odot}) \leq 11.5$.   
The percentage of red galaxies is a good indicator of quenching rate. It is well shown that, for the galaxies with more stellar mass, a larger fraction of galaxies has been quenched into quiescent.  
This seems to be consistent with the scenario of RS buildup (\citealt{Goncalves+12}) that massive star-forming galaxies were formed  and subsequently quenched into massive RS galaxies at earlier epoch.  
At later epochs star formation shifts to less massive galaxies, and then they are quenched into the red galaxies at the fainter end of the RS (``downsizing''; \citealt{Cowie+96, Noeske+07}).
Figure \ref{fig3} shows  a very low fraction ($\sim$ 10\%)  of blue galaxies with $M_*>10^{11}M_{\odot}$, implying that either quenching in these massive blue galaxies begins shortly after they reach these high masses, or that most blue galaxies begin to quench at a lower mass such that very few galaxies remain blue as they approach this mass regime. 

Moreover, the ascending trend of the green faction with mass can be seen at $z>1.5$.   
Assuming that blue star-forming galaxies start to be partially quenched into green galaxies, and ultimately quenched to be red and dead galaxies, 
Figure \ref{fig3} suggests that the most massive star-forming galaxies are rather unsustainable because $\sim$40\% of them  have been partially quenched into GV galaxies. 
Additionally, a remarkable redshift evolution can be found for  the blue fraction in Figure \ref{fig3}, indicating that a larger faction of blue galaxies tend to be quenched into green and red galaxies with the passage of time.  Our results are consistent with recent studies on the sSFR of the ``main sequence'' of star-forming galaxies,   finding that the average sSFR of the ``main sequence'' of star-forming galaxies is a mildly declining function of stellar mass (e.g., \citealt{Karim+11}). 
This implies that the bulk of star formation in more massive galaxies is completed earlier than that of lower-mass galaxies.

\subsection{Turnoff at $z\sim 2$ ?}

Some studies show that  star formation is quenched for an increasingly large fraction of the galaxies at $z<2$, which leads to a steep decline of  the globally averaged sSFR. 
\cite{M&D+14} study the SFH from UV and IR data and find that the star-formation rate density (SFRD) may peak at some point probably between $z = 2$ and 1.5, and followed by a exponential decline to the present day, with an \emph{e}-folding timescale of 3.9 Gyr.  
This cosmic SFRD picture can be further supported by our results on the properties of galaxies, which are in the process of being quenched since $z \sim 2$.
Firstly, green and blue galaxy population are found to have similar distributions of parametric measurements (i.e., $r_e$ and $n$) and nonparametric measurements (i.e., $C$, $G$, and $M_{20}$) at $z > 2$, and they come to exhibit different distributions since $z \sim 2$. 
Second, the AGN fraction for the green galaxy population at $z < 2$ is significantly higher than that of red and blue populations, whereas red galaxies at $z>2$ are found to have a higher AGN fraction, comparable to that of green galaxies at $z>2$.  
As one probable mechanism for truncating ongoing star formation, AGN activity seems to be more commonly found in completely and partially quenched galaxies, corresponding to  red and green populations, respectively. 
However, since $z \sim 2$, AGN activity tends to be more associated with the quenching process in green galaxies. 
Third, the distribution of local densities for green galaxy population comes to be significantly different with that of red galaxies since $z \sim 1.5$. 
It is therefore interesting that the estimated peak redshift of the cosmic SFRD roughly coincides with the transition in galaxy structural and AGN activity properties and that the preference for red galaxies to live in denser environments than those of green and blue galaxies emerges at $z<1.5$, subsequent to the peak in the cosmic SFRD. 

\section{Summary}
We present an analysis of massive galaxies with $\log M_*/M_{\odot} \geq 10$  at \textbf{$0.5 < z <2.5$} in all five CANDELS/3D-$HST$ fields, which are separated into red, green, and blue galaxy populations.
Our investigation focuses on properties of the three galaxy populations, including  dust attenuation, 
morphology, structural parameters, AGN fraction, and environmental density.
We redid the SED fitting ourselves to derive stellar masses and dust attenuation using the FAST code (\citealt{Kriek+09}) with \cite{Maraston+05} stellar population templates (which include the effects of AGB stars) and multiwavelength SEDs taken from 3D-$HST$ (\citealt{Skelton+14}). 
We compare our redshift-dependent definition of the BC, GV, and RS based on the extinction-corrected rest-frame $(U-V)$ color versus stellar mass diagram to the traditional $UVJ$ diagram-based separation of quiescent and star-forming galaxies (e.g., \citealt{Williams+09}). 
Structural evolution of galaxies are studied through statistics of three differnt standpoints: morphology classification,  parametric measurements (Sersic index $n$, half-light radius $r_e$, and compactness $\Sigma_{1.5}$), and nonparametric measurements (concentration $C$, Gini coefficient $G$, and second-order moment of the 20\% brightest pixels of a galaxy $M_{20}$).
Local surface density, $\Sigma_{10}$, is introduced as a local environment indicator to study the environmental effect on star formation quenching.
The fraction of AGN host galaxies is also derived to examine whether green galaxies have higher probabilities of hosting an AGN and to determine any possible interdependence between AGN feedback and star formation.

Our main conclusions can be summarized as follows:

1. The green galaxy population has intermediate distributions of structural parameters, such as $C$, $G$, $M_{20}$, and $n$, between red and blue galaxy populations at $z<2.0$. The green galaxy population seems to be a distinct transition population  when a star-forming galaxy is being quenched. A larger fraction of blue galaxies come to be quenched since $z \sim 2.0$. 

2. The correlation between morphology and galaxy population is quite significant. Green galaxies are found to have an intermediate distribution of morphological types between those of red and blue galaxies. Over cosmic time, a larger fraction of blue galaxies appear to be late-type galaxies, including the LTDs and IRRs, whereas a larger fraction of red galaxies are found to have striking bulges. 

3. The distributions of parametric and nonparametric measurements show that the growth of bulge accompanies the quenching process. 
The diversity of structure for three galaxy populations  become more significant from $z=2$ to present day.

4.  Green galaxies are found have the highest AGN  fraction ($\sim 20\%$ over $0.5<z<2.5$) . This suggests that AGN feedback plays an important role in star formation quenching. For the red galaxies at $z>2$,  a similarly higher AGN fraction suggests that AGN feedback may also help to maintain  the quiescence of galaxies. 

5. The three galaxy populations have the similar distributions of  environmental density at $z>1.5$.  
Since $z \sim 1.5$, the red galaxy population comes to exhibit a local density distribution that differ from those of the green and blue populations. 
A peak at the high-density end appears for the red galaxies with lower redshift ($0.5< z<1$), supporting the scenario that galaxy clusters form at $z \sim 1$.  

6. The fractions of the three galaxy populations as functions of mass support a ``downsizing'' quenching picture that the bulk of star formation in more massive galaxies is completed earlier than that of lower-mass galaxies.

\acknowledgments
This work is based on observations taken by the 3D-$HST$ Treasury Program (GO 12177 and 12328) with the NASA/ESA $HST$, which is operated by the Association of Universities for Research in Astronomy, Inc., under NASA contract NAS5-26555.
This work is supported by the National Natural Science Foundation of China (nos. 11673004, 11433005, 11173016) and by the Research Fund for the Doctoral Program of Higher Education of China (no. 20133207110006). We would like to thank Prof. Xu Kong, at the University of Science and Technology of China, and Prof. Xu Zhou, at the National Astronomical Observatories of China, for their valuable discussions.



\end{document}